# Analysis of DNA thermal stability across a broad range of thionine concentrations

Evgeniya Usenko, Alexander Glamazda*, Vladimir Valeev, Victor Karachevtsev

B. Verkin Institute for Low Temperature Physics and Engineering of the National Academy of Sciences of Ukraine, 47 Nauky Ave., Kharkiv 61103, Ukraine

* Corresponding author:

Dr. Alexander Glamazda,

B. Verkin Institute for Low Temperature Physics and Engineering of the National Academy of Sciences of Ukraine, 47 Nauky Ave., Kharkiv 61103, Ukraine

e-mail: glamazda@ilt.kharkov.ua

Evgeniya Usenko (0000-0002-8446-5504 – ORCID ID)

Alexander Glamazda (0000-0003-3048-8732 – ORCID ID)

Vladimir Valeev (0009-0008-6659-7280 – ORCID ID)

Victor Karachevtsev (0000-0003-4580-6465 – ORCID ID)

## Abstract

Interest in studying the interaction of small molecules with DNA is caused by the need to develop new, highly effective, and low-toxic drugs for cancer treatment. The strong and highly specific binding of thionine with DNA makes it a promising candidate for use in medicine and pharmacology. In this study, DNA-thionine complexes in aqueous solutions were investigated using UV-Vis absorption spectroscopy. The thermal stability of native DNA was studied in a broad range of thionine concentrations. The mechanisms of thionine binding to DNA, depending on the concentration of thionine, have been established. At low thionine concentrations ($[c_{th}] \leq 1.5$ mg/L), thionine molecules intercalate between the base pairs of the DNA double helix. At a thionine concentration of 1.5 – 10 mg/L, the groove binding and external electrostatic interaction of positively charged thionine with negatively charged biopolymer phosphate groups of the DNA backbones is preferable. In all cases, the interaction of thionine with DNA leads to an increase in the thermal stability of the polynucleotide. These findings provide valuable insight into the concentration-dependent molecular mechanisms of DNA-small molecule interactions, supporting the rational design of anticancer and antimicrobial agents, as well as exploiting molecular probes for nucleic acid detection, imaging, and other biomedical applications.

# 1 Introduction

The study of the molecular mechanisms of interaction of aromatic heterocyclic molecules with deoxyribonucleic acids has been of particular interest. Increased attention to the study of such systems is due, first of all, to the possibility of their use in the development of new chemotherapeutic agents, as well as methods for cancer treatment [1–4]. It's well known that flat small molecules are capable of binding to DNA through intercalation interaction [4,5]. Strong and highly specific intercalation binding of such molecules to genomic DNA leads to distortion to its structure and, thus, is capable of stopping the uncontrolled division of cancer cells [5]. Another non-invasive therapeutic action of such molecules may also include photosensitization, which can be used for targeted cleavage of the DNA backbone for photodynamic therapy of tumours and other diseases [6,7]. The intercalation binding properties of such molecules can also be used as diagnostic probes for studying the structure of DNA [8-10]. For instance, thionine, a phenothiazine dye, is an excellent candidate for DNA intercalation studies because its planar structure allows it to bind effectively between the base pairs of the DNA helix [11]. Thionine exists in the form depending on the pH of the medium: in neutral and slightly acidic solutions (pH ~ 4 – 7), thionine exists as a cation; in an alkaline medium (pH > 10), thionine can convert to the neutral leuco form due to deprotonation; in a highly acidic medium (pH < 2), the deprotonated form can form [12,13]. Thionine has also attracted increasing attention due to its potential use in semiconductors, as an energy sensitizer, as a probe for studying various microenvironments, including micelles and polymer matrices, in the preparation of functionalized nanocomposite materials, and photoelectrochemical cells with high quantum efficiency [14–19]. Recent studies have reported the application of thionine as a charge neutralizer for impedance-based DNA biosensors using its specific binding to both single- and double-stranded DNA [20]. In Ref. [21] the structural and energetic aspects of thionine binding to single- and double-stranded helical conformations of calf thymus DNA were determined using absorption, fluorescence, competitive dialysis, circular dichroism, and calorimetry. Spectroscopic data indicate that thionine binds preferentially to the double-stranded DNA. Thermodynamic parameters (enthalpy of binding, binding Gibbs' energy, entropic contribution to the binding) of binding obtained by isothermal titration calorimetry showed that, from the energetic point of view, thionine is the most favoured to bind to the double-stranded DNA conformation compared to the single-stranded one. The spectroscopic and calorimetric methods were used to study the thionine binding to three natural DNAs with different GC-pair content [4]. Results indicate that thionine primarily binds through intercalation to DNA with the highest GC-pair content and the weakest to DNA with the highest AT pair content. Further spectroscopic studies with synthetic double-stranded polynucleotides (poly(dG-dC)•poly(dG-dC), poly(dG)•poly(dC), poly(dA-dT)•poly(dA-dT), and poly(dA)•poly(dT)) confirmed that thionine binds to all tested sequences, with significantly stronger binding to GC-rich and alternating sequences than to AT-rich or homosequences [5].

The study [4] found that thionine binds to DNA in a non-cooperative manner, alters DNA conformation, and increases thermal stability. However, thermal denaturation data were reported for only one thionine concentration. Further research is needed to assess how temperature affects DNA stability across a wider range of thionine concentrations. This is especially important given evidence that thionine may be toxic to living cells [22–25].

The toxic effects of thionine on living organisms have been well-documented. Specifically, research has shown that thionine inactivates the frog sperm nucleus [23] and exerts a toxic influence on anaerobic glycolysis [24]. Furthermore, it has been observed to cause structural changes in rat mast cells and inhibit cellular metabolism, thereby preventing cell damage [25]. In addition to these harmful effects, a more recent study reported the unique ability of thionine to serve as a molecular adhesive by immobilizing proteins and DNA [26].

It is known that certain dyes bind to double-stranded polynucleotides through multiple mechanisms [27], with each mechanism becoming dominant within a specific P/D concentration range (where P/D is the molar polymer-to-dye ratios) [28]. Tuite and Kelly [29] were the first to propose, based on spectroscopic data, that thionine binds to DNA not only by intercalation but also through external electrostatic interaction (Fig. 1). While most research highlights intercalation as the primary and strongest binding mechanism [4,5,21], it is important to note the dual nature of intercalating molecules. They can play a positive role by blocking the proliferation of cancer cells or a negative one by disrupting the replication of healthy DNA. A fundamental step in DNA replication is the separation of the double strand into two single strands, a process driven by the helicase enzyme [30]. In a laboratory setting, this can be simulated through thermal denaturation, which serves as an analogue to helicase activity and allows for the study of DNA stability under various conditions. Since intercalation is a key mechanism for inhibiting cancer cell division, it is crucial to understand how temperature affects DNA stability across different thionine concentrations. Currently, this specific information is not available in the literature. Therefore, the present study aims to provide data on the effects of temperature and thionine concentration on DNA stability at pH7. The goal is to establish the temperature-concentration range in which a particular type of thionine binding to DNA is dominant and to determine the thermodynamic parameters of this interaction.

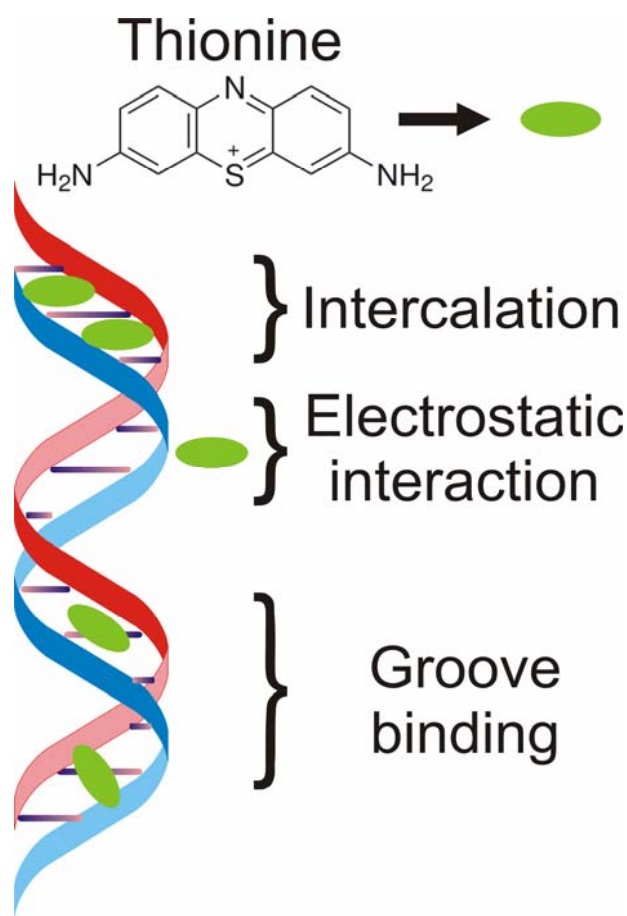

**Fig. 1** Chemical structure of thionine in cationic form. The different modes of thionine binding to DNA through intercalation, groove, and external binding

## 2 Materials and Methods

### 2.1 DNA-thionine complexes

The native DNA from the calf thymus manufactured by Serva (Germany) was used in the present work. Thionine was purchased from Sigma-Aldrich (Europe, Product number: 861340, $M_w$ = 287.34 g/mol). Thionine was dissolved in an aqueous solution. The calf thymus DNA was added to a buffer solution of $10^{-3}$ M sodium cacodylate $(CH_3)_2AsO_2Na \cdot 3H_2O$ from Serva (Germany), at pH7. Cacodylate buffer does not contain phosphate ions, which can compete with DNA phosphate groups for binding to cationic thionine. This competition can lead to an underestimation of the mechanisms of thionine bonding to DNA and distortion of thermodynamic parameters. Cacodylate buffer has a low ionization enthalpy and $pK_a$, with a pH close to physiological, which makes it ideal for thermodynamic studies and minimizing thermal effects unrelated to the thionine-DNA interaction itself, ensuring high measurement accuracy [31]. The low concentration of cacodylate buffer ($10^{-3}$ M) allowed us to neglect the formation of complexes of its anions with positively charged thionine molecules. The effect of thionine on the biopolymer structural conformation was studied by adding the required amount of thionine solution to the DNA buffer solution. The DNA phosphorus concentration [P] was $(6.9 \pm 0.6) \times 10^{-5}$ M, which was determined by the molar extinction coefficient ($\varepsilon_m$ = 6600 $M^{-1} \times cm^{-1}$) at the DNA absorption maximum ($\nu_m$ = 38,500 $cm^{-1}$).

### 2.2 UV-Vis absorption spectroscopy

UV-Vis and differential UV absorption spectra were recorded with a Specord UV-Vis spectrophotometer (Carl Zeiss, Jena, Germany). The extinction coefficient of a DNA's double helix structure depends on the mutual angle orientation of the intrinsic electric dipole moments of stacked base pairs. The light absorption will be suppressed (hypochromism) if the moments are aligned. Upon disordering the moments placed in the basis plane, the light absorption will increase (hyperchromism).

The complexation of different ligands with DNA may induce changes in the biopolymer structure, and hence it may lead to a change in the optical absorption spectra. The slight change in the optical absorption spectra can be revealed with differential UV (DUV) absorption spectroscopy utilizing the four-cuvette scheme of the signal comparison [32]. DUV spectra were normalized to the concentration of polynucleotide phosphates as follows: $\Delta\varepsilon(\nu)$ = $\Delta A(\nu)/[P]$, where $\Delta\varepsilon(\nu)$ is the change in light absorption at $T = T_0 = 20$ °C and $\Delta A(\nu)$ is the change in the optical density of the solution at $T = T_0$.

### 2.3 Thermal studies of DNA

The thermal evaluation of DNA structural stability sheds light on the nature of ligand binding to DNA. Upon heating, the ordered structure of the biopolymer is disrupted due to the breaking of the hydrogen bonds between the complementary nitrogenous base pairs. The temperature at which half of the total complementary nitrogen base pairs separate is called the melting temperature ($T_m$), and it characterizes the structural stability of the biopolymer.

In the present work, the thermal studies of DNA stability in the absence and presence of thionine were performed at $\nu_m$ = 38,500 см$^{-1}$, which corresponds to the maximum absorption of DNA. The registration of the absorption intensity was carried out using the double-cuvette scheme [32].

To ensure accuracy, each spectral and thermal measurement was repeated a minimum of three times. The resulting measurement error was maintained at 1 – 2%.

## 3 Results

### 3.1 Thionine influence on the thermal stability of biopolymer

The melting curves of DNA in the absence and presence of thionine are presented in Fig. 2. It is evident from this figure that the melting curve of the pure DNA has a typical S-shape. Upon heating, absorption hyperchromism (h > 0) is observed, caused by the helix-coil transition.

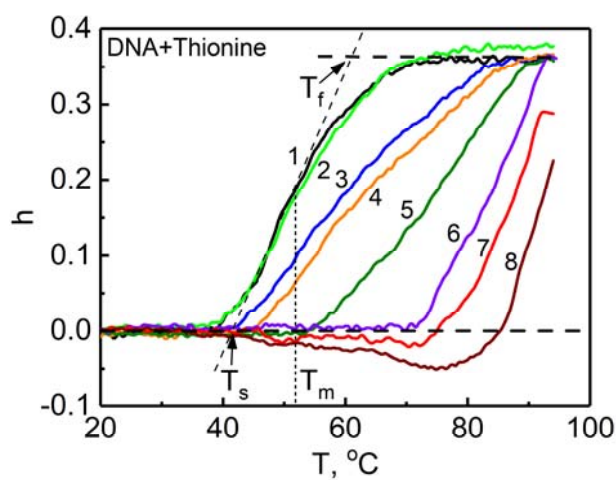

**Fig. 2** Temperature dependence of hyperchromic coefficient (h) of DNA without (curve 1) and with (curves 2 − 8) thionine: "1" – $[c_{th}]$ = 0 mg/L; "2" – $[c_{th}]$ = 0.1 mg/L; "3" – $[c_{th}]$ = 1 mg/L; "4" – $[c_{th}]$ = 1.5 mg/L; "5" –$[c_{th}]$ = 2.5 mg/L, "6" – $[c_{th}]$ = 5 mg/L; "7" – $[c_{th}]$ = 7.5 mg/L; "8" – $[c_{th}]$ = 10 mg/L

The addition of thionine (0.1 mg/L) to the DNA solution does not disrupt the double-stranded structure and conformation of DNA, since the melting curve in this case practically coincides with the melting curve for pure DNA. However, the injection of $[c_{th}]$ = 1 mg/L induces a change in the shape of the melting curve: DNA melting becomes less cooperative, and it occurs in a wider temperature interval (Fig. 2). The melting interval was determined as follows: $\Delta T = T_f - T_s$, where $T_s$ and $T_f$ are the temperatures at which DNA melting starts and finishes, respectively (Fig. 2). That was found by the intersection of the tangent to the melting curve at point $T = T_m$ with the lines corresponding to the condition $dh/dT = 0$. The melting interval values derived from the analysis of Fig. 2 are presented in Fig. 3. An increase in thionine concentration leads to a broadening of the melting interval, which peaks near 1.5 mg/L and is followed by a subsequent narrowing. A possible reason for the increase in the melting range is the intercalation of thionine molecules into the biopolymer. The intercalation of thionine between stacking base pairs does not, by itself, break the hydrogen bonds between complementary base pairs. However, it does distort the DNA structure by increasing the spacing between stacking base pairs. The intercalation leads to an elongation of the double helix and a weakening of the stacking interactions between adjacent base pairs. This distortion of the DNA structure can weaken the existing hydrogen bonds between complementary base pairs, which, in turn, can lead to localized breaking of these bonds, especially under increased temperature or mechanical stress. Furthermore, intercalation causes a localized unwinding of the double helix. This creates regions of reduced stability in the DNA, allowing these regions to melt independently at different temperatures. All these factors collectively result in a more protracted melting process and a wider melting interval. At the same time, at [$c_{th}$]≥1.5 mg/L, the DNA melting range narrows (Fig. 3). A possible reason for this effect is the involvement of other types of thionine binding to DNA (groove binding and external electrostatic interaction of thionine with the phosphate backbone of DNA). Continuing the analysis of the melting curves presented in Fig. 2, it can be noted that when 2.5 mg/L thionine is added, a longer non-cooperative "tail" appears on the melting curve in the temperature range of 20 – 54 °C, which indicates an increase in the length of single-stranded regions at the ends of the biopolymer. The number of single-stranded regions increases with the growth of the thionine concentration in the DNA solution, as evidenced by the increase in the length of the non-cooperative "tail". At the same time, at [$c_{th}$] ≥ 1.5 mg/L, the DNA melting range narrows (Fig. 3). Intercalation is known to distort the DNA structure, which weakens the hydrogen bonds between complementary base pairs and makes them more easily susceptible to breaking, particularly upon heating. This explains why, at concentrations exceeding 1.5 mg/L, single-stranded regions form more readily at the ends of the DNA helix upon heating. As the thionine concentration increases, these single-stranded regions become progressively longer. Also, it can be noted that in the temperature range of 45 – 75 °C at [$c_{th}$] ≥ 7.5 mg/L, hypochromism is observed, which is replaced by hyperchromism, upon further heating. A possible reason for the revealed hypochromism may be the formation of the conglomerates consisting of several DNA-thionine complexes [33].

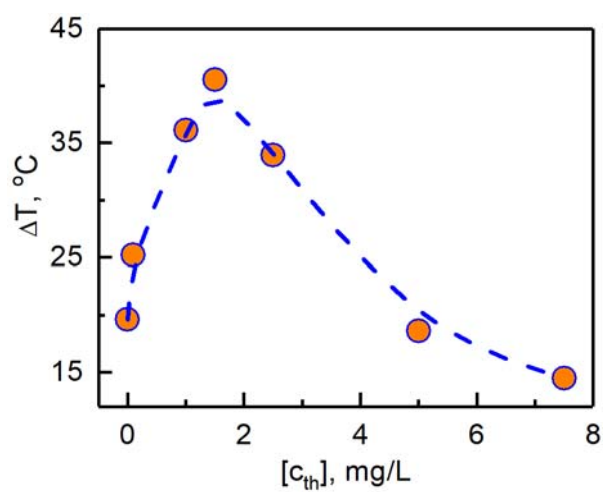

**Fig. 3** Melting interval (ΔT) of DNA in the DNA-thionine suspension. Data were interpolated with a spline function

The analysis of data presented in Fig. 2 made it possible to determine the $T_m$([$c_{th}$]) and $h$([$c_{th}$]) dependences (Figs. 4a and 4b). Fig. 4a clearly shows that injecting thionine into the DNA solution raises the DNA melting temperature by 34.7 °C at concentration up to 7.5 mg/L. Moreover, as [$c_{th}$] increases, the slope of the $T_m$ dependence changes: in the range [$c_{th}$] = 0 – 2.5 mg/L, $dT_m/d[c_{th}]$ = 8.56 deg/mg; at [$c_{th}$] = 2.5 – 5 mg/L, $dT_m/d[c_{th}]$ = 3.72 deg/mg; at [$c_{th}$] = 5 – 7.5 mg/L, $dT_m/d[c_{th}]$ = 1.11 deg/mg. The pronounced drop in $dT_m/d[c_{th}]$ when [$c_{th}$] > 2.5 mg/L can be explained by an increase in the length of unwound single-stranded regions at the ends of double-stranded DNA. Thionine can also bind to these single-stranded regions. This interaction leads to changes in the melting temperature that differ from thionine's intercalation with DNA. Therefore, thionine's binding to single-stranded DNA regions directly affects $T_m$([$c_{th}$]), resulting in distinct temperature changes across concentration ranges.

As for the change in the $h$ value with the concentration, the injection of thionine up to [$c_{th}$] = 1.5 mg/L has negligible effect on this parameter, with fluctuations within ±3% (Fig. 4b). With a subsequent increase in the dye concentration, a gradual decrease in h is observed, and at a thionine concentration of $[c_{th}] = 7.5$ mg/L, h decreases by 16%. This means that at $[c_{th}] = 1.5$ mg/L or below, DNA maintains its predominantly double-stranded conformation. At higher thionine concentrations, the single-stranded regions appear in native DNA. Increasing the number of these unwound regions reduces the h value.

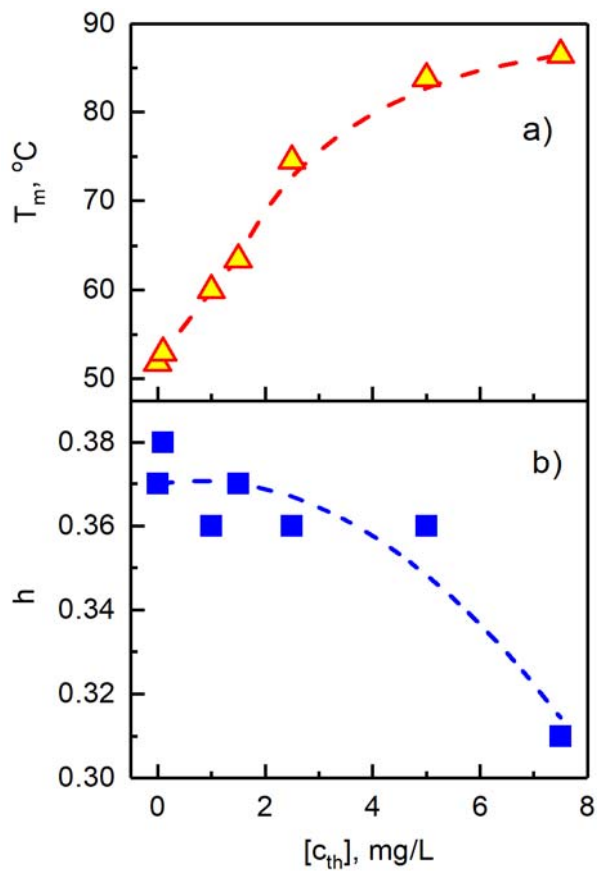

**Fig. 4** a) Effect of thionine concentration on the melting temperature $T_m([c_{th}])$ and b) hypochromic coefficient $h([c_{th}])$ of DNA. Data were interpolated with a spline function

The maximum value of h(T) achieved with complete strand separation, $(h_m)_i$, can be used as a measure of DNA helicity at $T = T_0$. Without thionine, $(h_m)_0$ of DNA is 0.37, which corresponds to the degree of helicity $\Theta_{T_0} = 1$ (or 100%). Thus, the degree of helicity of DNA can be calculated as: $\Theta = (h_m)_i/(h_m)_0 \times 100\%$, where $(h_m)_i$ is the maximum value of the hyperchromic coefficient of thermally denatured DNA at $\nu_m = 38,500$ cm$^{-1}$ ($\lambda = 260$ nm). The indices "0" and "i" determine the absence and presence of thionine. The dependence of the degree of DNA helicity on thionine concentration is presented in Table 1. The increase in $[c_{th}]$ up to 1.5 mg/L has little effect on the DNA helicity (within ±3%). Thus, at $[c_{th}] = 1.5$ mg/L, the degree of DNA helicity is 97 – 100%. However, further increase in the thionine content (5 mg/L $\leq [c_{th}] \leq$ 7.5 mg/L) reduces $\Theta$, and at $[c_{th}] = 7.5$ mg/L the helicity drops to 84%. This result correlates with the observation that non-cooperative sites appear on the melting curve at thionine concentrations exceeding 1.5 mg/L, corresponding to the melting of single-stranded DNA fragments in the double helix. These regions are formed for the following reasons. As is known, hydrogen bonds are preserved during intercalation, but the double helix structure is distorted. This distortion can lead to a weakening of the hydrogen bonds and their easier breaking, particularly upon heating. This explains why, at concentrations exceeding 1.5 mg/L, single-stranded regions form more easily at the ends when the DNA is heated. As the thionine concentration increases, these regions become longer. At thionine concentrations below 1.5 mg/L, even if hydrogen bonds are broken upon heating, they are able to reform, thus preserving the double helix. Furthermore, the $\Delta T([c_{th}])$ dependence shows a maximum at 1.5 mg/L. This observation indicates that at concentrations exceeding 1.5 mg/L, the equilibrium shifts toward the implementation of alternative types of interaction between thionine and DNA: moving away from intercalation and towards the external electrostatic interaction of thionine with the phosphate groups of both double- and single-stranded DNA, as well as binding in the minor groove of DNA.

**Table 1**

Dependence of the degree of DNA helicity ($\Theta$) on the concentration of thionine ($[c_{th}]$) and P/D.

| $[c_{th}]$, mg/L | 0 | 0.1 | 1 | 1.5 | 2.5 | 5 | 7.5 |
|---|---|---|---|---|---|---|---|
| P/D | - | 189 | 19 | 14 | 8 | 4 | 3 |
| $\Theta$, % | 100 | 103 | 97 | 100 | 97 | 97 | 84 |

Temperature measurements revealed that during the formation of DNA-thionine complexes, there is a threshold concentration at which one type of binding is replaced by another. The spectral analysis, detailed in the next section, is therefore crucial to confirm and elucidate the binding mechanisms of DNA-thionine complex formation.

**3.2 Some spectral features of thionine bounded with DNA**

Fig. 5a shows the absorption spectrum of an aqueous solution of thionine. The obtained spectrum is in good agreement with the absorption spectrum of thionine presented in Ref. 29. It can be seen that there are two most intense peaks: at about $\lambda_{max1}$ = 283 and $\lambda_{max2}$ = 600 nm. The absorption band at 283 nm is corresponding to the $\pi-\pi^*$ transition of the phenothiazine ring, whereas the band at 600 nm is due to the n$-\pi^*$ transitions of the C=N bond [34]. The 566 nm shoulder can be associated with dimers [12].

The DUV absorption spectra of thionine in complex with DNA at different thionine concentrations are presented in Fig. 5b. There are two well-identified minima (at about $\lambda_{min1}$ = 283 nm and $\lambda_{min2}$ = 593 nm for $[c_{th}]$ = 1 mg/L) and two maxima. For both minima, an increase in thionine concentration results in hypochromism of absorption accompanied by a bathochromic shift of the isosbestic point at about 611 nm. The observed hypochromism of the main $\lambda_{min1}$ and $\lambda_{min2}$ bands is caused by the binding of thionine to DNA [35]. The evolution of the DUV spectra suggests that the main mechanism of thionine binding to DNA is dye intercalation [4]. Similar spectral changes were observed when studying DNA complexes with phenothiazine derivatives, including methylene blue and acridine orange, and included hypochromism and a bathochromic shift in the absorption spectra upon binding of organic dyes to DNA [35,36].

The band at $\lambda_{max3}$ (at about 522 nm) appears at a high thionine concentration ($[c_{th}] \geq 20$ mg/L) and grows with concentration. In addition, hyperchromism of the absorption of the long-wavelength maximum at $\lambda_{max4}$ is observed in the DUV spectra of the DNA-thionine complex within a thionine concentration range from 1 to 7.5 mg/L (Fig. 6). With a further increase in the $[c_{th}]$ value, the band at $\lambda_{max4}$ shifts to the long-wavelength region and its intensity plateaus (Fig. 6). We believe that in our measurements, thionine remains predominantly in the monomeric form at most concentrations, as evidenced by the shape of DUV spectra with a pronounced band at $\lambda_{max2}$ (Fig. 5b). At thionine concentrations of 20 mg/L and 30 mg/L, the formation of complex thionine aggregates on the DNA surface is possible, indirectly indicated by the appearance of a band at $\lambda_{min3}$. The appearance of this band can be attributed to the formation of the thionine dimers (H-type), consistent with similar spectral changes observed for thionine in the presence of gold nanoparticles [37]. The analysis of data presented in Fig. 6 shows that the normalized absorption intensity of the band at $\lambda_{max4}$ increases with the concentration up to 7.5 mg/L and then reaches a plateau. We believe that this growth is caused by the thionine intercalation between nucleobases, which ends at 7.5 mg/L. Thus, the binding mode of thionine with DNA changes above 7.5 mg/L, suggesting minor groove binding and external electrostatic binding of thionine to the DNA phosphate backbone. It is clear that at certain thionine concentrations, both intercalation and groove binding can occur. The obtained value of the threshold concentration of thionine, at which one type of binding to DNA is replaced by another, is somewhat different from the value that was obtained in temperature studies in the previous section of the present work. The obtained threshold concentration of thionine, at which one type of binding is replaced by another, exceeds the value obtained in the temperature-dependent studies described in the previous chapter. With increasing temperature, the intercalation process proceeds significantly faster than at room temperature. This is caused by the increased flexibility of DNA with increasing temperature, which makes dye intercalation more likely, even at low concentrations [38].

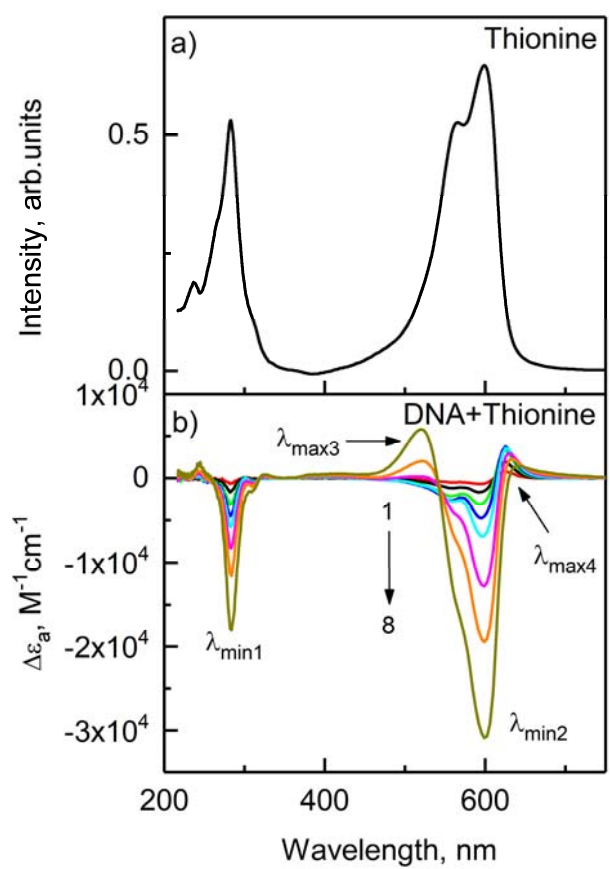

**Fig. 5** a) Absorption spectrum of an aqueous solution of thionine ($[c_{th}]$ = 0.12 g/L) and b) DUV spectra of DNA-thionine complexes with $[c_{th}]$ = 1, 2.5, 5, 7.5, 10, 15, 20, and 30 mg/L (curves 1-8). All experimental data were obtained at room temperature

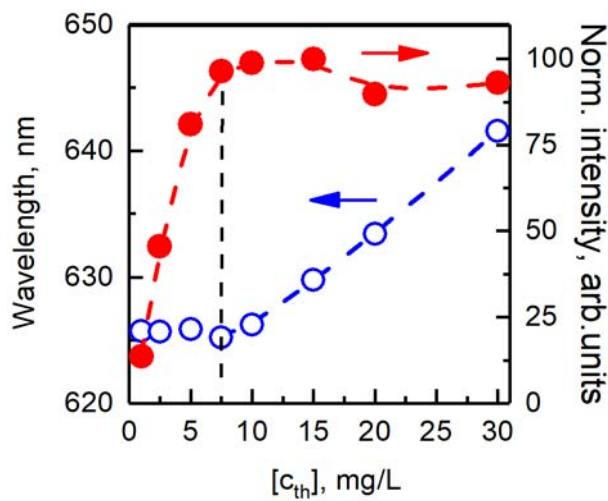

**Fig. 6** Evolution of the peak position and normalized absorption intensity of the band at λ$_{max4}$ extracted from the data presented in Fig. 5b as a function of [c$_{th}$]. Data were interpolated with a spline function

The performed analysis of data presented in Fig. 5b allows us to plot the concentration dependences of the intensity of the most intense specific absorption bands (λ$_{min1}$ and λ$_{min2}$) of DNA-thionine solution as thionine concentration increases (Fig. 7). It is evident from Fig. 7 that both concentration dependencies demonstrate a decrease in absorption extinction of DNA-thionine solution over the entire concentration range of thionine. Both dependencies exhibit a similar evolution up to a thionine concentration of 7.5 mg/L. Above [c$_{th}$] = 7.5 mg/L, the $\Delta\varepsilon_a$([c$_{th}$]) dependencies diverge. The reason for such spectral features is most likely the stronger influence of both the negatively charged phosphate backbone and the polarizing effect of the aqueous environment on the electron density localized near the heteroatoms involved in the groove binding and external electrostatic interaction. In contrast, during intercalation, the dye molecule's electronic structure is surrounded by an extended π-conjugated electronic system within a less polar environment inside DNA, which facilitates partial delocalization of the electron density. Thus, similar effects on the π−π* and n−π* transitions are observed during intercalation.

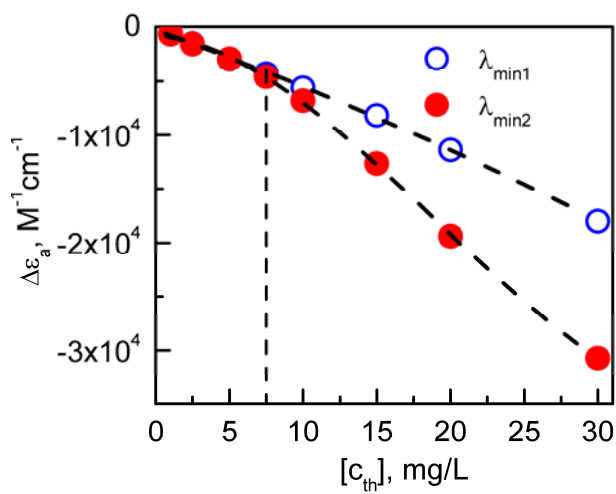

**Fig. 7** Dependence of the absorption intensity of two characteristic bands at about $\lambda_{min1}$ and $\lambda_{min2}$ revealed in absorption spectra of DNA-thionine as a function of thionine concentration

# 4 Discussion

The data obtained in this study indicate that, at thionine concentration of 0.1 – 1.5 mg/L during heating, the dominant interaction is thionine intercalation into the DNA double helix. This interaction results in a stabilizing effect and increases in the melting temperature. Indeed, according to the available experimental [29] and theoretical [11] studies, the primary type of interaction of thionine with DNA is intercalation. This interaction is considered to be the strongest and should be realized first. Similar intercalation mechanisms are observed for related compounds such as methylene blue, acridine orange, ethidium bromide, and SYBR Green [10,36,39-41]. As in the case of the interaction of thionine with DNA, the intercalation of methylene blue molecules between DNA base pairs stabilizes the native structure of DNA, as evidenced by an increased melting temperature of the complexes compared to pure DNA [39,40]. Furthermore, thionine exhibits a higher affinity for GC pairs than for AT pairs, as demonstrated by binding constants reported for DNA with varying GC content [4,5]. This is evidenced by the values of the binding constants obtained in the work [4] for DNA with different GC pair contents. The binding constant of thionine with double-stranded DNA is ~ $(2 – 4) \times 10^5$ M$^{-1}$, and this value increases with higher GC content. Other types of interaction of thionine with DNA (the groove binding and external electrostatic interaction) are realized in the range of $[c_{th}]$ = 1.5 – 10 mg/L [21,29]. These types of interactions between thionine and DNA also stabilize the DNA double helix, as evidenced by the data in Fig. 4a. This figure clearly shows that the melting temperature increases across the entire concentration range tested. The external electrostatic interaction is classified as weak and occurs between the negatively charged phosphate groups of DNA and the positively charged atoms of thionine. For single-stranded DNA, the binding constant is about an order of magnitude lower, approximately $10^4$ M$^{-1}$ [21]. Therefore, during external interaction, thionine predominantly binds to double-stranded DNA but can also interact with single-stranded regions.

Several types of thionine binding to DNA are confirmed by the type of concentration dependence for the melting interval (Fig. 3). It can be seen that this dependence has a bell-shaped form: at $[c_{th}]$ = 1.5 mg/L, a broadening of the DNA melting interval is observed compared to pure DNA. This effect is due to the intensive implementation of the intercalation type of thionine interaction with DNA. Then, the binding sites are saturated, and the double-stranded regions become inaccessible for the intercalation of thionine molecules. However, there are binding sites available for the implementation of other binding modes, in particular, the groove binding and the external electrostatic interaction of thionine with double- and single-stranded DNA. The manifestation of these binding modes leads to the formation of a maximum on the $\Delta T([c_{th}])$ dependence, which begins to decrease with a further increase in the thionine concentration, and thus, this leads to a narrowing of the melting interval. According to the data of the work [42,43], it is this bell-shaped dependence that indicates the presence of several types of ligand binding to DNA. If there were only one type of binding, this dependence should reach a plateau.

According to the data obtained in this work, it can be concluded that with increasing temperature, the external electrostatic interaction and groove binding begin to be realized at a significantly lower concentration of thionine (≥ 1.5 mg/L). This result seems quite natural, since, according to the available experimental data [33], temperature enhances the process of formation of single-stranded regions in double-stranded DNA. This leads to a more intensive formation of intercalated DNA and, accordingly, a shift in equilibrium towards the realization of other types of thionine binding to DNA.

In addition, Fig. 4a shows that all types of interaction are accompanied by a stabilizing effect. The thermodynamic reason for these effects is the same in both cases. It is that the binding constants of thionine to the double helix are higher than to the single helix. In the case of binding to phosphates, this difference is due to the lower charge density on the surface of the single-stranded polymer compared to that in the case of the double helix. In the case of intercalation, it is due to the lower free energy of binding of thionine to N1 of guanine and N3 of cytosine in single-stranded DNA compared to the energy of formation of GN1-thionine-N3C complexes in the double helix [21].

Since temperature measurements identified a threshold concentration causing a shift in the DNA-thionine binding mode, the spectral analysis of these complexes performed at room temperature is crucial. This analysis can be key to both confirming and explaining the detailed binding mechanisms involved in the formation of the DNA-thionine complex. The analysis of room-temperature DUV absorption spectra of the thionine-DNA complex suggests that thionine binds to DNA primarily via intercalation up to a threshold concentration. As thionine concentration increases, the main absorption minima (at about $\lambda_{min1}$ = 283 nm and $\lambda_{min2}$ = 593 nm revealed at $[c_{th}]$ = 1 mg/L) show hypochromism and a bathochromic shift, which are characteristic of dye binding and similar to other phenothiazine derivatives. This hypochromism is a key indicator of intercalation, which is thought to be complete at about 7.5 mg/L. Above this concentration, the binding mode shifts from intercalation to the minor groove binding or external electrostatic binding to the DNA phosphate backbone, although both modes may coexist. At very high concentrations ($[c_{th}] \geq 20$ mg/L), the appearance of new bands suggests the formation of thionine aggregates (possibly H-type dimers) on the DNA surface [37]. The observed threshold concentration for the binding mode change is slightly higher than that found in temperature-dependent studies, where increased DNA flexibility at higher temperatures facilitates intercalation even at lower concentrations.

## 5 Conclusion

This study employed UV-visible spectroscopy and DNA thermal denaturation to investigate the effects of temperature and thionine concentration on DNA stability in aqueous solutions at pH7. Thionine interaction with DNA was found to increase the melting temperature of DNA across all tested thionine concentrations of 0.1 − 7.5 mg/L. Specific concentration ranges were identified in which a particular mode of thionine binding to DNA predominates. There is a concentration at which the interaction type shifts from intercalation to the groove binding and electrostatic interaction. The results obtained in the present work can help shed light on the molecular mechanisms of small molecule binding to double-stranded DNA. That will stimulate the development of new highly effective, and most importantly, less toxic drugs for the treatment of viral and cancer diseases.

**Author contribution** E.U.: data collection, data analysis and interpretation, and drafting the article. A.G.: data analysis and interpretation, conception/design of the work, critical revision of the article. V.V.: data collection and data analysis. V.K.: drafting the article, resources, and supervision.

**Funding** The authors acknowledge partial financial support from National Academy of Sciences of Ukraine (Grant №0123U100628). E.U. acknowledges Wolfgang Pauli Institute, Vienna, Austria for the financial support (Pauli Postdoc research training scholarship in the field of Data analysis in molecular biophysics in the context of the WPI thematic program "Numerical models in Biology and Medicine" (2023/2024). A.G. acknowledges Wolfgang Pauli Institute, Vienna, Austria for the financial support (Pauli Postdoc research training scholarship in the field of Data analysis in